\documentstyle[times,aps,eqsecnum,prb]{revtex}
\input epsf.sty
\begin{document}
\wideabs{
\title{Near-field EM wave scattering from random self-affine fractal
             metal surfaces: spectral dependence of local field enhancements
             and their statistics in connection with SERS}

\author{Jos\'e A. S{\'a}nchez-Gil and Jos\'e V. Garc{\'\i}a-Ramos}
\address{Instituto de Estructura de la Materia,
Consejo Superior de Investigaciones Cient{\'\i}ficas,
Serrano 121, E-28006 Madrid, Spain
}
\author{Eugenio R. M{\'e}ndez}
\address{Divisi{\'o}n de F{\'\i}sica Aplicada, Centro de
Investigaci{\'o}n Cient{\'\i}fica y de Educaci{\'o}n Superior de
Ensenada \\ Ensenada, Baja California 22800 M{\'e}xico}
\date{submitted to \prb{}}
\maketitle

\begin{abstract}

By means of rigorous numerical simulation calculations based on the
Green's theorem integral equation formulation, we study the near EM
field in the vicinity of very rough, one-dimensional self-affine
fractal surfaces of Ag, Au, and Cu (for both vacuum and water
propagating media) illuminated by a $p$ polarized field. Strongly
localized enhanced optical excitations ({\it hot spots}) are found,
with electric field intensity enhancements of close to 4 orders of
magnitude the incident one, and widths below a tenth of the incoming
wavelength. These effects are produced by roughness-induced
surface-plasmon polariton excitation.  We study the characteristics of
these optical excitations as well as other properties of the surface
electromagnetic field, such as its statistics (probability density
function, average and fluctuations), and their dependence on the
excitation spectrum (in the visible and near infrared).  Our study is
relevant to the use of such self-affine fractals as surface-enhanced
Raman scattering substrates, where large local and average field
enhancements are desired.

\end{abstract}
\pacs{}
}

\section{Introduction}

Recently, single molecule probing by means of Surface Enhanced Raman
Scattering (SERS) has been reported both on Ag single nanoparticles
\cite{nie97} and on Ag colloidal aggregates;\cite{kneipp97} the
latter exploits the extremely large near infrared (NIR) Raman
scattering cross sections of dye molecules.\cite{kneipp96} Bearing in
mind how inefficient normal spontaneous Raman scattering is,
enhancement factors of $10^{14}$ or larger are required to achieve
single molecule detection.\cite{nie97}

Typically, SERS is known to yield Raman signals enhanced by a factor
${\cal G}\sim 10^6$ with respect to those of conventional Raman
scattering.\cite{sers_mosk,sers_wok,sers_otto,sers_aroca,sharep} Two
mechanisms are responsible for such enhancement factors: the
surface-roughness-induced intensification of the electromagnetic (EM)
field both at the pump frequency and at the Raman-shifted frequency
(EM mechanism), and the charge-transfer mechanism.  The former
mechanism is widely accepted to be the most relevant one from the
quantitative standpoint, providing gains of ${\cal G}_{EM}>10^4$ in
most experimental configurations. Extensive theoretical work has been
devoted to the explanation of the EM mechanism (cf. e.g. the reviews
Refs.~\onlinecite{sers_mosk,sers_wok,sharep}), and the consensus is
that what underlies such EM field enhancement (FE) factors is the
roughness-induced excitation of surface-plasmon polaritons \cite{spp}
(SPP), either propagating along a continuous surface (extended SPP),
confined within metal particles (particle plasmon resonances), or even
confined due to Anderson localization (localized SPP).

In light of the SERS enhancement factors estimated for single molecule
detection it is evident that, in addition to the well known average
SERS enhancement factors, extremely large EM fields must appear in the
vicinity of SERS substrates, even if the charge-transfer mechanism is
also known to be especially intense, as in Ref.~\onlinecite{kneipp96}.
This is supported by the observation, through photon scanning tunneling
microscopy (PSTM), of very intense and narrow EM modes (called {\it
hot spots}) on rough metal surfaces \cite{tsai94,bozh95,zhang98} or
rough metal-dielectric films.\cite{gres99}  Interestingly, these
rough metal surfaces used as SERS substrates posses, in some cases,
scaling properties within a sufficiently wide range of scales
(physical fractality). The substrates can present self-similarity, as
the widely employed colloidal aggregates,
\cite{tsai94,jull87,forna87,sii88,stock92,jcis94} or self-affinity,
as in the case of deposited colloids or evaporated or etched rough
surfaces and thin films.\cite{zhang98,chen95,douke95,bara}

Therefore, inasmuch as the quantitative evaluation of the surface EM
field is central to the SERS effect, knowledge of the EM scattering
process for surface models as realistic as possible is obviously
needed. In recent years, the theoretical efforts have been directed
towards either describing through approximate methods realistic
surface models,\cite{sharep,tsai94,gres99,xu93,sha96,sha98} or using
the full EM theory to study simplistic surface models,
\cite{pendry96,oc97,jcp98} though introducing increasingly complex
properties.\cite{oc97,jcp98}

In this paper we study the near EM field scattered in the vicinity of
rough, one-dimensional self-affine fractal surfaces of Ag, Au, and Cu,
with the aim of determining the appearance of strong local optical
excitations ({\it hot spots}) and characterizing them with regard to
their spatial and spectral width, their polarization, and their
excitation spectra; in addition to that, the global optical response
of such fractal surfaces will be studied through the statistical
properties of the surface EM fields. Both local and global responses
are discussed in light of the influence on SERS.  For this purpose, we
make use of numerical simulation calculations based on the Green's
theorem integral equation formulation,\cite{ann,nieto,josaa91}
rigorous from the classical EM standpoint. Unlike recent work also for
self-affine fractals (though sensitively less rougher) based strictly
on magnetic field calculations,\cite{oc97,jcp98} the magnitude that
naturally arises in this formulation when applied to 1D surfaces and
$p$ polarization (the one relevant for its light-SPP coupling
selectivity), we fully characterize here the surface and near electric
field components (crucial in SERS and other nonlinear optical effects)
through simple expressions in terms of the magnetic field and its
normal derivative. The necessary details of the theoretical
formulation are given in Sec.~\ref{sec_sform}. The local optical
excitations are studied in Sec.~\ref{sec_loc} and their statistical
properties in Sec.~\ref{sec_sta}, leaving for Sec.~\ref{sec_con} the
conclusions of this work.

\section{SCATTERING FORMULATION}
\label{sec_sform}

\subsection{Surface integral equations}
\label{sec_sform_sie}

\begin{figure}
\epsfxsize=3.25in \epsfbox{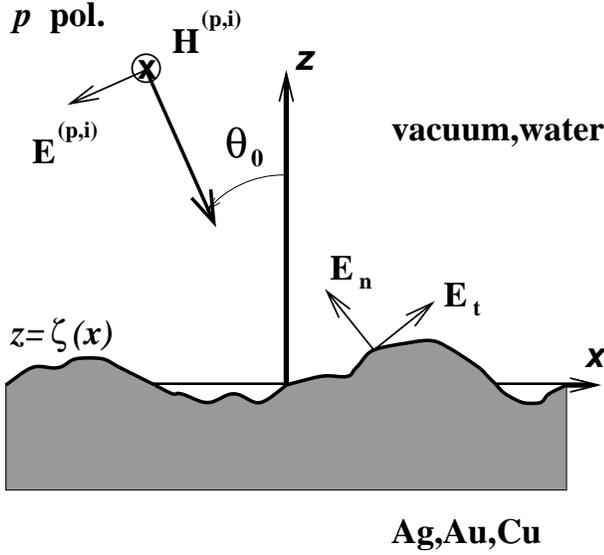}
\caption{Schematic of the scattering geometry with the electromagnetic field
        vectors for $p$ linear polarization.}
\label{fig_sca}
\end{figure}

The scattering geometry is depicted in Fig.~\ref{fig_sca}. A rough
metal surface $z=\zeta(x)$ is the substrate onto which molecules are
adsorbed in SERS typical experimental configurations. The
semi-infinite metal occupying the lower half-space [$z\leq\zeta(x)$]
is characterized by an isotropic, homogeneous, frequency-dependent
dielectric function $\epsilon^<(\omega)$.  From the medium of
incidence, characterized by a frequency-dependent dielectric function
$\epsilon^>(\omega)$, a monochromatic, linearly polarized incident
beam of frequency $\omega$ impinges on the interface at an angle
$\theta_0$, measured counterclockwise with respect to the positive $z$
axis.  The polarization is defined as shown in Fig.~\ref{fig_sca}: the
magnetic (respectively, electric) field is perpendicular to the $xz$
plane for $p$ (respectively, $s$) polarization, also known as the
transverse magnetic (respectively, transverse electric) one.

We restrict the analysis to 1D surfaces (invariant along the $y$
direction), without any loss of generality as far as the physics
underlying the SERS EM mechanism is concerned.
\cite{pendry96,jcp98} It has been shown \cite{ann,nieto,josaa91} that
the latter condition simplifies considerably the formulation based on
the integral equations resulting from the application of Green's
second integral theorem (with the help of the Sommerfeld radiation
condition). In such circumstances, the starting 3D vectorial problem
can be cast into a 2D scalar one, where the unknown is the
$y$-component of either the magnetic field [$H^{(p)}_y({\bf
r},\omega)$, with ${\bf r}\equiv (x,z)$] for $p$ polarization, or the
electric field [$E^{(s)}_y({\bf r},\omega)$] for $s$ polarization; no
depolarization takes place when pure $s$ or $p$ polarized fields are
incident. This simplification is very convenient from the analytical
and numerical point of view, and yields straightforwardly the far
field scattered intensity.\cite{ann,nieto,josaa91} We point out that,
if the electric (respectively, magnetic) field is needed for $p$
(respectively, $s$) polarization, it can be calculated on the basis of
Maxwell equations, as we shall see below.

Let us focus on the electric field calculation for the case of $p$
polarization. Evidently this is the most relevant one for our SERS
problem (only $p$ polarized light can excite in the present
configuration the SPP responsible for the EM field enhancements), and
also to other interesting problems such as second harmonic generation
on metal surfaces \cite{sharep,shg} or PSTM studies.
\cite{tsai94,bozh95,zhang98} As mentioned above, the integral
equation formulation is simplified if written in terms of the magnetic
field amplitude. Our monochromatic incident field of frequency
$\omega$ is a Gaussian beam of half-width $W$ in the form:\cite{ann}
\begin{mathletters}\begin{eqnarray}
  \lefteqn{H_y^{(p,i)}(x,z|\omega) =\exp\left\{\imath k_{\epsilon}
    (x\sin\theta_0-z\cos\theta_0)\right.}  \nonumber \\ && \left.
    \times[1+w(x,z)]\right\}
    \exp\left[- \frac{(x\cos\theta_0+z\sin\theta_0)^2}{W^{2}}\right] , \\
  \lefteqn{w(x,z) =\frac{1}{k_{\epsilon}^2 W^2}
    \left[\frac{2}{W^{2}}(x\cos\theta_0 +z\sin\theta_0)^2-1\right]} ,
\end{eqnarray}\label{eq_if}\end{mathletters}
where $k_{\epsilon}=n_c^> \omega/c$ and $n_c^>=\sqrt{\epsilon^>}$.
From now on, since a time harmonic dependence $e^{-\imath\omega t}$ is
assumed, the functional dependence on frequency will be omitted unless
necessary for the sake of clarity. The surface integral equations that
fully describe the EM linear scattering problem for $p$ polarization,
in the geometry of Fig.~\ref{fig_sca}, are
\begin{mathletters}\begin{eqnarray}
 \lefteqn{H_y^{(p,i)}({\bf r})+\frac{1}{4\pi}\int^{\infty}_{-\infty}
    \!\!\! \gamma 'dx'\left[ H_y^{(p,>)}({\bf r'})
    \frac{\partial G^>({\bf r},{\bf r'})}{\partial n'}\right. }
    \nonumber \\ & \left.  - G^>({\bf r},{\bf r'})
    {\displaystyle\frac{\partial H_y^{(p,>)}({\bf r'})}{\partial n'}}\right] &
    =  H_y^{(p,>)}({\bf r}), z>\zeta(x)
    \label{eq_ieuf} \\  &&=0, z<\zeta(x), \\
 \lefteqn{-\frac{1}{4\pi}\,\int^{\infty}_{-\infty}\! \gamma 'dx'\left[
    H_y^{(p,<)}({\bf r'})\frac{\partial G^<({\bf r},{\bf r'})}{\partial n'}
    \right. }  \nonumber \\ &\left.
     -G^<({\bf r},{\bf r'}) {\displaystyle\frac{\partial
    H_y^{(p,<)}({\bf r'})}{\partial n'}}\right] &
     = 0,  z>\zeta(x)  \label{eq_iebc}\\ &&
     = H_y^{(p,<)}({\bf r}), z<\zeta(x), \label{eq_ielf}
\end{eqnarray}\label{eq_ie}\end{mathletters}
where $H_y^{(p,>)}({\bf r})$ and $H_y^{(p,<)}({\bf r})$ are the
magnetic field in the upper ($z>\zeta$) and lower ($z<\zeta$)
semi-infinite half-spaces, and the normal derivative is defined as
$\partial / \partial n\equiv (\hat{\bf n}\cdot\nabla)$, with $\hat{\bf
n}\equiv\gamma^{-1} (-\zeta'(x),0,1)$ and
$\gamma=(1+(\zeta'(x))^2)^{1/2}$.  The 2D Green's function $G$ is
given by the zeroth-order Hankel function of the first kind
$H_0^{(1)}$.

The four integral equations (\ref{eq_ie}) fully describe the
scattering problem for $p$ polarization in terms of the $y$ component
of the magnetic field. Analogous integral equations can be obtained
for $s$ polarization dealing with the $y$ component of the electric
field. In order to solve for the surface field and its normal
derivative, defined as the functions $H(x)$ and $\gamma^{-1} L(x)$,
two of the integral equations (note that they are not independent),
typically Eqs.~(\ref{eq_ieuf}) and (\ref{eq_iebc}), are used as
extended boundary conditions, leading to two coupled integral
equations once one invokes the continuity conditions across the
interface:
\begin{mathletters}\begin{eqnarray}
 H(x)= &&H_y^{(>)}({\bf r})\mid_{z=\zeta^{(+)}(x)} =
   H_y^{<)}({\bf r})\mid_{z=\zeta^{(-)}(x)}, \\
 \gamma^{-1} L(x)= &&\left[ \frac{\partial H_y^{(>)}
   ({\bf r})}{\partial n}\right] _{z=\zeta^{(+)}(x)}  \nonumber \\
  & &  =\frac{\epsilon^>}{\epsilon^<}\left[ \frac{\partial H_y^{(<)}
   ({\bf r})}{\partial n}\right] _{z=\zeta^{(-)}(x)},
\end{eqnarray}\label{eq_cc}\end{mathletters}
with $\zeta^{(\pm)}(x)=\lim_{\varepsilon\rightarrow 0}(\zeta(x)\pm
\varepsilon)$. The resulting system of integral equations can be
numerically solved upon converting it into a system of linear
equations through a quadrature scheme,\cite{josaa91} the unknowns
being $H(x)$ and $L(x)$. Then Eqs.~(\ref{eq_ieuf}) and (\ref{eq_ielf})
permit to calculate the scattered magnetic field in the upper incident
medium and inside the metal, respectively.

But what if the magnitude of interest is the electric field? This is
indeed the situation in SERS where the surface electric field locally
excites the molecule vibrations that produce the Raman-shifted
radiation that is detected. In Refs.~\onlinecite{oc97,jcp98}, the EM
field enhancement factor has been defined as the normalized magnetic
field intensity:
\begin{eqnarray} \sigma_{H}(\omega) & = & \frac{H_y^{(p)}}{\mid\!
H_y^{(p,i)}\!\mid^2}. \label{eq_feh} 
\end{eqnarray} 
Even if the enhancement factor thus defined closely resembles the
correct total electric field enhancement factor, we are evidently
losing information about the different electric field components, in
turn relevant to the SERS polarization selectivity.

\subsection{$p$ polarization: Electric field}
\label{sec_sform_ef}

In order to obtain the electric field components from the $y$
component of the magnetic field, use can be made of the Maxwell
equation
\begin{equation}
\nabla\times{\bf H}=-\imath\frac{\omega}{c}\epsilon{\bf E}.
\label{maxwell}
\end{equation}
In the incident medium, Eq.~(\ref{eq_ieuf}) provides the only non-zero
component of the magnetic field. Use of Maxwell's equation
(\ref{maxwell}) leads to the following electric field components:
\begin{mathletters}\begin{eqnarray}
 \lefteqn{E_x^{(p,>)}({\bf r})=E_x^{(p,i)}({\bf r}) -\imath
 \frac{c}{4\pi\omega\epsilon^>}\int^{\infty}_{-\infty}\!\!\! \gamma 'dx'
 \left[ H_y^{(p,>)}({\bf r'}) \frac{}{} \right. } \nonumber \\
 && \left.\times\frac{\partial^2 G^>
    ({\bf r},{\bf r'})}{\partial z\partial n'} - \frac{\partial
    G^>({\bf r},{\bf r'})}{\partial z}
    \frac{\partial H_y^{(p,>)}({\bf r'})}{\partial n'}\right] \\
 \lefteqn{E_y^{(p,>)}({\bf r})=0} \\
 \lefteqn{E_z^{(p,>)}({\bf r})=E_z^{(p,i)}({\bf r}) +\imath
 \frac{c}{4\pi\omega\epsilon^>}\int^{\infty}_{-\infty}\!\!\! \gamma 'dx'
 \left[ H_y^{(p,>)}({\bf r'}) \frac{}{} \right. } \nonumber \\
 && \left.\times\frac{\partial^2 G^>
    ({\bf r},{\bf r'})}{\partial x\partial n'} -\frac{\partial
    G^>({\bf r},{\bf r'})}{\partial x}
    \frac{\partial H_y^{(p,>)}({\bf r'})}{\partial n'}\right]
\end{eqnarray}\label{eq_ef}\end{mathletters}
These equations can be rewritten in terms of the source functions
$H(x)$ and $L(x)$ as follows:
\begin{mathletters}\begin{eqnarray}
 \lefteqn{E_x^{(p,>)}({\bf r})=E_x^{(p,i)}({\bf r}) - \frac{\omega}{4c}
  \int^{\infty}_{-\infty}\!\!\!\gamma 'dx'
  \left\{ H(x')\right.} \nonumber\\
 && \times\left[\frac{z-\zeta(x')}{\mid {\bf r}-{\bf r'}\mid^2}
    ({\bf n}\cdot({\bf r}-{\bf r'}))
    H_2^{(1)}(k_{\epsilon}\mid {\bf r}-{\bf r'}\mid) \right.\nonumber \\
    && \left. -\frac{1}{\gamma' k_{\epsilon}\mid {\bf r}-{\bf r'}\mid}
    H_1^{(1)}(k_{\epsilon}\mid {\bf r}-{\bf r'}\mid)\right] \nonumber \\
    && \left.-L(x')\frac{z-\zeta(x')}{\gamma' k_{\epsilon}\mid {\bf r}-{\bf r'}\mid}
    H_1^{(1)}(k_{\epsilon}\mid {\bf r}-{\bf r'}\mid)\right\} \\
 \lefteqn{E_y^{(p,>)}({\bf r})=0} \\
 \lefteqn{E_z^{(p,>)}({\bf r})=E_z^{(p,i)}({\bf r}) - \frac{\omega}{4c}
  \int^{\infty}_{-\infty}\!\!\!\gamma 'dx'
  \left\{ H(x')\right.} \nonumber\\
 && \times\left[-\frac{x-x'}{\mid {\bf r}-{\bf r'}\mid^2}
    ({\bf n}\cdot({\bf r}-{\bf r'}))
    H_2^{(1)}(k_{\epsilon}\mid {\bf r}-{\bf r'}\mid) \right.\nonumber \\
    && \left. -\frac{\zeta'(x')}{\gamma' k_{\epsilon}\mid {\bf r}-{\bf r'}\mid}
    H_1^{(1)}(k_{\epsilon}\mid {\bf r}-{\bf r'}\mid)\right] \nonumber \\
    && \left. + L(x')\frac{x-x'}{\gamma' k_{\epsilon}\mid {\bf r}-{\bf r'}\mid}
    H_1^{(1)}(k_{\epsilon}\mid {\bf r}-{\bf r'}\mid)\right\}
\end{eqnarray}\label{eq_efhl}\end{mathletters}
where the explicit form of the Green's function has been taken into
account, leading to the appearance of 1st and 2nd order Hankel
functions of the first kind $H_1^{(1)},H_2^{(1)}$. For the Gaussian
incident field given by Eq.~(\ref{eq_if}), the electric field
components are
\begin{mathletters}\begin{eqnarray}
 E_x^{(p,i)}({\bf r})= && \frac{\imath}{n_c^>} H_y^{(p,i)}({\bf r})
  \left[ \frac{}{} \imath\cos\theta_0 (1+w(x,z))\right. \nonumber \\
  && -\left(\imath\frac{4}{k_{\epsilon}^2 W^4}
  (x\sin\theta_0-z\cos\theta_0)\right.
   \nonumber \\ && \left. \left.
 -\frac{2}{k_{\epsilon} W^2}\right)\sin\theta_0(x\cos\theta_0+z\sin\theta_0)\right]\\
 E_y^{(p,i)}({\bf r})= && 0 \\
 E_z^{(p,i)}({\bf r})= && \frac{\imath}{n_c^>}
  H_y^{(p,i)}({\bf r}) \left[ \imath\sin\theta_0 (1+w(x,z))\right. \nonumber \\
  && +\left(\imath\frac{4}{k_{\epsilon}^2 W^4}
  (x\sin\theta_0-z\cos\theta_0)\right.
   \nonumber \\ && \left. \left.
  -\frac{2}{k_{\epsilon} W^2}\right)\cos\theta_0(x\cos\theta_0+z\sin\theta_0)
  \right]
\end{eqnarray}\label{eq_ief}\end{mathletters}

Equations (\ref{eq_efhl}) and (\ref{eq_ief}) provide the electric
field components in the incident medium of the resulting $p$-polarized
EM field, incident plus scattered from the rough surface. The
scattered electric field involves an additional surface integral in
terms of the source functions, previously obtained (numerically) from
the above mentioned coupled integral equations. Analogous expressions,
not shown here, for the corresponding electric field components inside
the metal can be obtained from Eq.  (\ref{eq_ielf}). On the other
hand, recall that a similar procedure can be straightforwardly
developed to yield the magnetic field components in the case of
$s$-polarized EM waves as surface integrals in terms of the surface
electric field ($y$ component) and its normal derivative.

\subsection{$p$ polarization: Normal and tangential surface electric
field}
\label{sec_sform_sef}

It should be pointed out that when trying to evaluate the electric
field on the surface, or even very close to it, from Eqs.\
(\ref{eq_efhl}), non-integrable singularities appear associated with
the Green's functions derivatives for vanishing arguments. Use of
expressions (\ref{eq_efhl}) for the evaluation of the electric field
close to the surface will produce unphysical results. To deal properly
with this situation, more care should have been taken in doing the
derivatives of the integral (\ref{eq_ieuf}) describing the magnetic
field, whose integrand already exhibits singularities, though
integrable.\cite{ann,nieto,josaa91} A simple way to work around this
problem consists of evaluating the electric field at the surface
itself, and we have found very simple relations connecting the normal
and tangential components of the electric field (see
Fig.~\ref{fig_sca}) with the surface magnetic field and its normal
derivative [cf. Eqs.~(\ref{eq_cc})]. These are
\begin{mathletters}\begin{eqnarray}
 E^{(p,>)}_n(x)& =& \frac{\imath c}{\omega\epsilon^>} \gamma^{-1}
   \frac{dH(x)}{dx} \\
 E^{(p,>)}_t(x)& =& -\frac{\imath c}{\omega\epsilon^>}
   \gamma^{-1} L(x).
\end{eqnarray}\label{eq_sef}\end{mathletters}
These expressions are extremely useful, for they facilitate
considerably our study of the SERS EM mechanism.

Thus, taking advantage of the one-dimensional scattering geometry
(which, although is not general, is not inappropriate to study the
SERS EM mechanism \cite{pendry96,jcp98}), we only have to deal with
the y-component of the magnetic field to obtain a simplified,
basically exact solution to the scattering problem from the classical
EM viewpoint, at the excitation frequency. The numerical solution of
the resulting integral equations yields the surface magnetic field and
its normal derivative as the main results. The drawback of working
with the magnetic field, when the quantity of interest is the electric
field, is avoided by expressions (\ref{eq_sef}), which allows us to
obtain the surface electric field with the only additional algebra of
calculating a spatial derivative.

We now properly define the electric field enhancement factors for
either a single component or the total field as the normalized
intensities:
\begin{eqnarray}
 \sigma_{\alpha}(\omega) & = & \frac{\mid\! E^{(p,>)}_{\alpha}\!\mid^2}
    {\mid\! E^{(p,i)}\!\mid^2}, \label{eq_fea} \\
 \sigma(\omega) & = & \frac{\mid\! E^{(p,>)}\!\mid^2}{\mid\!
    E^{(p,i)}\!\mid^2}, \label{eq_fe}
\end{eqnarray}
with $\alpha=n,t,x,z$ and $E=\mid {\bf E}\mid^2=E_n^2+E_t^2=E_x^2+E_z^2$.

\subsection{Numerical implementation}
\label{sec_sform_ni}

The numerical procedure has been implicitly outlined above; further
details have been given in Ref.~\onlinecite{jcp98}.  Self-affine random
fractal surfaces numerically generated by means of Voss' fractional
Brownian motion algorithm \cite{feder,voss} are studied. This kind of
fractals exhibit self-affine scaling properties in a broad spatial
range,\cite{jcp98} and have properties that resemble those of some
SERS substrates, such as cold-deposited metal films or etched metal
surfaces.\cite{douke95,bara} In the numerical calculations, surface
realizations of length $L=10.29 \mu$m, consisting of $N_p=n_i N$
sampling points obtained by introducing $n_i=4,6,8$ or 10
cubic-splined interpolating points into a sequence of $N=201$ points
extracted from each generated fractal profile with $N_f=1024$ points;
note the considerably larger sample density with respect to that of
Ref.~\onlinecite{jcp98}. The statistical properties of the physical
quantities of interest will be calculated on the basis of Monte Carlo
simulations for an ensemble of fractal realizations.

\section{Local Field Enhancement: {\it Hot Spots}}
\label{sec_loc}

We now turn to the investigation of the occurrence of very large near
EM field enhancements. Particularly, we will concentrate on
self-affine fractals with Hurst exponent $H=0.1$ (namely, local
fractal dimension $D_f=2-H=1.9$), which have been shown in
Ref.~\onlinecite{jcp98} to give rise to large surface magnetic
fields. The lower scale cutoff $\xi_L\sim 50$ nm has been chosen to
resemble that of SERS substrates.\cite{jcis94,douke95} The upper
scale cutoff, typically $\xi_L\sim 50 \mu$m, is considerably larger
than the illuminated area $L$, and this is in turn sufficiently (in
order to avoid finite length effects) larger than the incoming
wavelength (0.4 $\mu$m$<\lambda <1.3 \mu$m). Thus physical scaling is
meaningful for the relevant interval of this scattering problem. The
effect of further reducing the lower scale cutoff will be investigated
elsewhere;\cite{pre_fe} in this regard, it should be recalled that
the minimum scale relevant to the far field pattern has been studied
for Koch fractals.\cite{ao97}

\subsection{Near Field Intensity Maps}
\label{sec_loc_hs}

\begin{figure}
\epsfxsize=3.25in \epsfbox{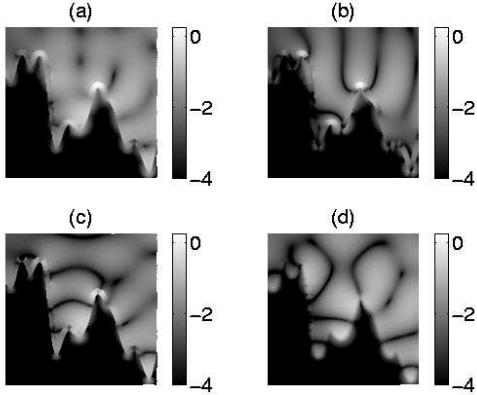}
\caption{Near field intensity images (in a $\log_{10}$ scale) resulting
        from the $p$-polarized scattering with $\theta_0=0^{\circ}$,
        $\lambda=514.5$ nm, and $W=L/4\cos\theta_0$, from a Ag fractal surface
        with $D=1.9$, $\delta=514.5$ nm, $L=10.29 \mu$m,  and $N_p=2000$.
        The area shown is 1$\times 1 \mu$m$^2$.
        (a) Electric field; (b) Electric field, $z$ component (vertical); (c)
        Electric field, $x$ component (horizontal); (d) Magnetic field.}
\label{fig_nfw1}
\end{figure}

In Fig.~\ref{fig_nfw1}, the intensity (on a logarithmic scale) of the
electric and magnetic near field in the vicinity of a self-affine Ag
surface with $D=1.9$ and rms height $\delta=514.5$ nm, in a particular
region (of about $1 \times 1 \mu$m$^2$), is shown for normal incidence
with light of wavelength $\lambda=514.5 \mu$m; in addition, the
intensities of the two different components ($x$ and $z$) of the
electric field are separately shown. Before analyzing the results,
some comments are in order with regard to the numerical
calculations. Whereas the electric and magnetic fields in vacuum away
from the interface are given by the integral equations (\ref{eq_efhl})
and (\ref{eq_ieuf}), respectively, and similarly for the EM field
inside silver, their corresponding values on the interface are
directly obtained from the source functions through Eqs.~(\ref{eq_cc})
and (\ref{eq_sef}). As mentioned in the preceding section,
Eqs.~(\ref{eq_ieuf}) and (\ref{eq_efhl}) exhibit singularities upon
approaching the surface, so that they are not accurate at points very
close to the surface, typically within distances smaller than the
surface sampling interval. Thus the EM field intensity in
Fig.~\ref{fig_nfw1} at distances from the surface smaller than
$4L/N_p$ for the magnetic field and $6L/N_p$ for the electric field
are obtained from the weighted values on the two closest sampling
points on the surface, taking explicitly into account for points
inside silver the continuity conditions for the magnetic
[Eq.~(\ref{eq_cc})] and electric field components. Despite that, some
slight (not inherent) mismatch might still appear when entering into
the surface field area, mostly in the intensities of the electric
field components.

It is, of course, expected that such continuity conditions across the
interface could be roughly observed in the calculations even if we
were not considering the above explicit matching. On the one hand, the
continuity of the tangential component of the magnetic field is neatly
seen in Fig.~\ref{fig_nfw1}(d); on the other hand, the continuity of
the tangential electric field is appreciable in Fig.~\ref{fig_nfw1}(b)
and (c) through the continuity of the $x$ (respectively, $z$)
component of the electric field at locally flat (respectively,
vertical) parts of the rough surface, whereas the discontinuity of the
normal component of the electric field (continuity of the normal
component of the displacement vector) is inferred from the
discontinuity of the $x$ (respectively, $z$) component of the electric
field at locally vertical (respectively, flat) areas.  Incidentally,
note also that the EM field inside silver decays very rapidly as
expected from the Ag skin depth $d=(c/\omega)
(-\epsilon^<)^{-1/2}\approx 27$ nm (cf. Ref.~\onlinecite{palik} for
the Ag dielectric constant).

\begin{figure}
\epsfxsize=3.in \epsfbox{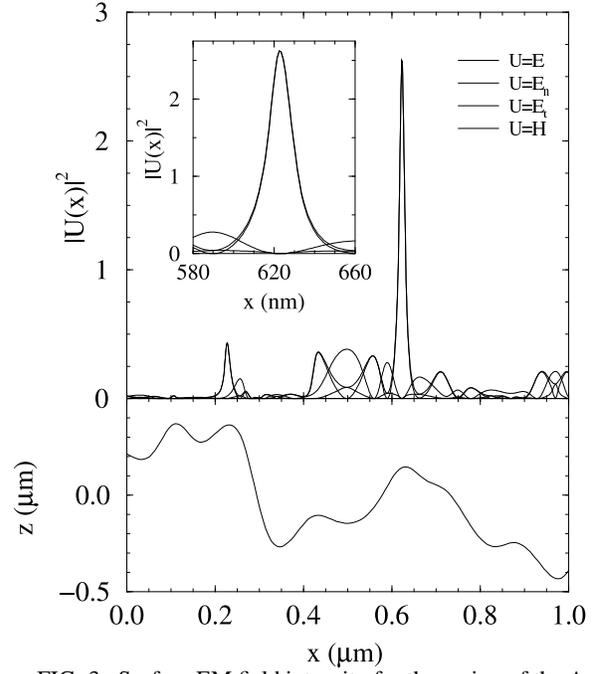}
\caption{Surface EM field intensity for the region of the Ag fractal 
        profile (also depicted) shown in Fig.~\protect{\ref{fig_nfw1}}. 
        Solid curve: Electric field; dashed curve: Electric field, normal 
        component; long-dashed curve: Electric field, tangential component; 
        dot-dashed curve: Magnetic field. The inset zooms in the largest 
        {\it hot spot}.}
\label{fig_sfw1}
\end{figure}

\begin{figure}
\epsfxsize=3.in \epsfbox{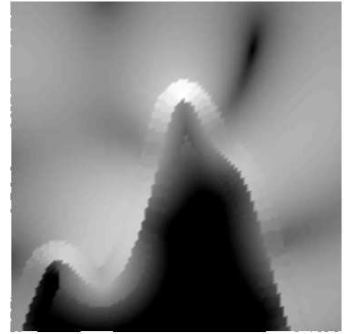}
\caption{Near electric field intensity  image (in a $\log_{10}$ scale)
        for the {\it hot spot} area shown in Fig.~\protect{\ref{fig_nfw1}}
        for a Ag fractal surface with $\theta_0=0^{\circ}$, $D=1.9$,
        $\delta=514.5$ nm, $\lambda=514.5$ nm, $L=10.29 \mu$m, $W=L/4\cos
        \theta_0$, and $N_p=2000$. The area shown is 500$\times 500$ nm$^2$.}
\label{fig_nfw1_hs}
\end{figure}

It is evident from Fig.~\ref{fig_nfw1} that the maximum local EM
fields are located right on top of the Ag surface, whereupon some
particularly bright spots appear. Thus we next plot in
Fig.~\ref{fig_sfw1} the intensities of the surface EM fields
(including electric tangential and normal components) for the surface
area shown in Fig.~\ref{fig_nfw1}, including the surface profile.
Very narrow peaks surrounded by dark areas are observed, whose widths
are well below half wavelength of the SPP ($\lambda_{SPP}/2=243$
nm). Some of these peaks can be considered as optical excitations
({\it hot spots}) where very large local FE's occur, such that non
linear optical processes would be strongly enhanced
therein.\cite{sharep} In particular, the largest in
Fig.~\ref{fig_sfw1}, shown in the inset, is of the order of
$\sigma\approx 6\cdot 10^2$. Note that our calculations identify the
local electric field component that is responsible for such FE: the
normal component. In Fig.~\ref{fig_nfw1_hs}, the near electric field
in the vicinity of this hot spot (zooming in Fig.~\ref{fig_nfw1}(a))
is given; interestingly, it is associated with a surface peak.

\begin{figure}
\epsfxsize=3.25in \epsfbox{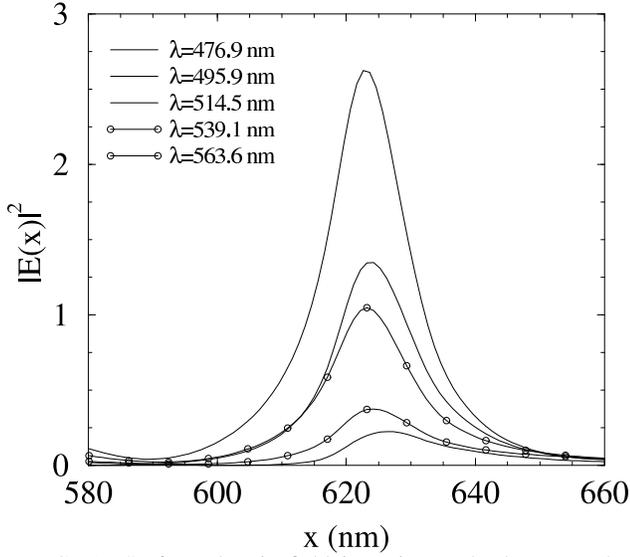}
\caption{Surface  electric field intensity at the hot spot shown in
         Fig.~\protect{\ref{fig_sfw1}} but for additional, slightly shifted
         incoming wavelengths $\lambda=$476.9, 495.9, 539.1, and 563.6 nm.}
\label{fig_sfw1+-}
\end{figure}

It has been experimentally shown by near-field microscopy that such
optical excitations rapidly disappear upon changing the frequency of
the incident radiation.\cite{zhang98} Our rigorous calculations
corroborate those experimental observations, as seen in
Fig.~\ref{fig_sfw1+-}, where the surface electric field intensity is
plotted for several incident wavelengths close to $\lambda=514.5$
nm. For the wavelengths $\lambda=$495.9 and 539.1 nm, the high
intensity spot is still clearly visible, though less bright. It fades
away, however, for larger frequency shifts, becoming barely visible
for $\lambda>563.6$ nm or $\lambda<476.9$ nm (about 10$\%$ frequency
shift).

The hot spot shown in Fig.~\ref{fig_nfw1_hs} is strongly polarized
along the normal to the surface. This is extremely relevant to SERS
spectroscopy, since it might impose selection rules to the vibrational
modes of the adsorbed molecule. Is it possible to find hot spots with
different polarizations? Only in the case of silver at wavelengths
close to the surface plasma wavelength, we have found certain spots
strongly polarized along the tangential direction too, though weaker
than those normally polarized. These tangentially polarized hot spots
exhibit FE factors not larger than $\sigma_t\approx 10^2$, and are
typically located in regions presenting larger values of $\sigma_n$:
this will be discussed elsewhere.\cite{pre_fe}

The occurrence of local optical excitations has been studied for the
same self-affine surface profile illuminated with different excitation
wavelengths, and also for metals such as Au and Cu. Although not shown
here, similar normally polarized hot spots are found on a broad
spectral range on all the self-affine surfaces of Ag, Au, and Cu. We
now discuss some of the characteristics of these hot spots.

It has been argued \cite{bozh95} that these hot spots are due to
Anderson localization of SPP. Theoretical works based on a dipolar
model, also demonstrate the possibility of creating strongly confined
and intense excitations on self-affine fractal surfaces;\cite{sha98}
in fact, in the case of random metal-dielectric films, Anderson
localization of surface plasmon modes is predicted.\cite{gres99} Our
numerical calculations, not subject to dipolar (and quasi-static)
restrictions, indeed reveal the existence of these kind of optical
excitations and, although compatible with the possibility of them
being due to Anderson localization of SPP, do not permit to draw
further conclusions in this respect. Our scattering geometry involving
the interaction between a propagating beam of light and a metal
surface does not lend itself well for the characterization of the SPP
Anderson localization phenomenon. To that end, the study of the
propagation and transmission of SPP through rough surfaces would be
more adequate.\cite{prb97} Only the fact that light can couple into
these, possibly localized, SPP modes through the roughness can be
inferred from our calculations and from the typical PSTM
configurations.\cite{bozh95}

It is also worth pointing out that the extremely rapid decay and
widening of the optical excitations in the near field (see
Figs.~\ref{fig_nfw1} and \ref{fig_nfw1_hs}) implies that PSTM images
taken at certain distance from the surface, leaving aside the rounding
effects of the tip, will manifest themselves as much wider and weaker
optical excitations, and this seems to be the case.
\cite{bozh95,zhang98} Direct probing of hot spots, on the other hand,
could be carried out by the nonlinear effects of physi- or
chemi-sorbed molecules.\cite{sharep,sha98}

\subsection{Spectral dependence}
\label{sec_loc_w}

\begin{figure}
\epsfxsize=3.25in \epsfbox{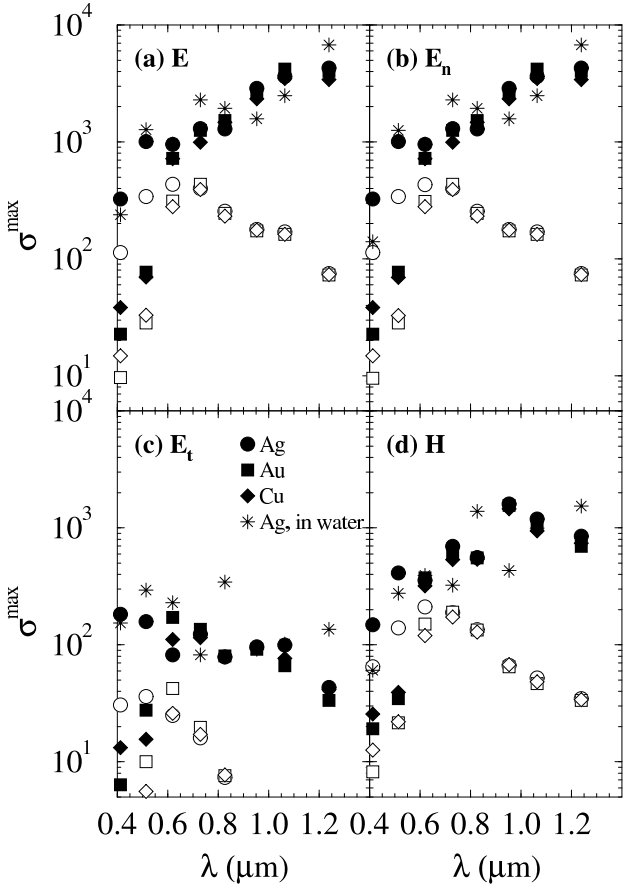}
\caption{Spectral dependence of the maximum local FE $\sigma^{max}$
        resulting from the $p$-polarized scattering with
        $\theta_0=0^{\circ},5^{\circ},10^{\circ},\ldots ,50^{\circ}$,
        and $W=L/4\cos\theta_0$, from fractal metal
        surfaces with $D=1.9$, consisting of $N_r=60$
        realizations of $L=10.29 \mu$m  and $N_p=1600$.
        (a) Electric field; (b) Electric field, normal component; (c)
        Electric field, tangential component; (d) Magnetic field.
        Circles, Ag; Squares, Au; Triangles, Cu.
        Filled symbols: $\delta=514.5$ nm; Hollow symbols: $\delta=102.9$ nm.
        Stars: water/Ag, $\delta=514.5$ nm.}
\label{fig_mfe_w}
\end{figure}

In order to analyze the polarization and spectral dependencies of the
optical excitations, we present in Fig.~\ref{fig_mfe_w} the maximum
local FE values found at Ag, Au, and Cu self-affine surfaces with
$D=1.9$ and $\delta=514.5$ nm, obtained from numerical calculations of
the surface EM field for an ensemble of $N_r=60$ realizations
generated as mentioned above (only the data from the central half of
each realization are used).  The results for weaker self-affine
surfaces ($\delta=102.9$ nm) used in Ref.~\onlinecite{jcp98} to compute
magnetic FE's are also shown. In addition, the case of having water as
incident medium (solvent) has been analyzed, though in
Fig.~\ref{fig_mfe_w} only the results for H$_2$O/Ag are shown.
Several remarks are in order with regard to Fig.~\ref{fig_mfe_w}.

Very large FE's appear for a wide spectral range covering the visible
and entering into the NIR. In the red and NIR parts of the spectrum,
all three metals being studied behave similarly, giving rise to hot
spots exhibiting strongly enhanced electric field intensities normal
to the surface, the tangential component tending to vanish. This
behavior can be understood in accordance with the spectral evolution
of the Ag, Au, and Cu dielectric constants,\cite{palik} all showing
increasingly large negative real parts, tending to the perfectly
conducting limit $\epsilon\rightarrow -\infty$ that predicts vanishing
tangential electric fields. For wavelengths $\lambda\gg 1240$ nm,
however, FE's are expected to slowly decrease as the surface is
``seen'' by the incoming radiation of increasing wavelength as
increasingly flatter. Other calculations, not shown here, indicate
that this is the case for $\lambda>2\mu$m. In fact, this decay is
observed at lower wavelengths (within the spectral interval covered by
Fig.~\ref{fig_mfe_w}) for the fractal surface with $\delta=102.9$ nm.

The optical responses of Au and Cu manifest significant differences
with respect to that of Ag in the blue part of the spectrum. The onset
of interband transitions, which takes place in Ag at $\lambda\approx
300$ nm unlike in Au and Cu (slightly below 600 nm), makes the
difference inasmuch as such transitions constitute a strong absorption
mechanism. Consequently, FE's should be significantly reduced for
wavelengths below the onset threshold, as is evident in
Fig.~\ref{fig_mfe_w} for Au and Cu below $\lambda\approx 600$ nm, but
not seen for Ag since the lower wavelength considered in
Fig.~\ref{fig_mfe_w} is above the Ag onset threshold.  Moreover,
silver surfaces at small incoming wavelengths approaching the surface
plasmon wavelength (but above the onset of interband transitions)
present strong local optical excitations tangentially polarized, as
mentioned above. These tangential-electric hot spots can lead to local
FE nearly comparable to those corresponding to the normal-electric
ones at such wavelengths, although more than an order of magnitude
weaker than those obtained at larger wavelengths [see
Fig.~\ref{fig_mfe_w}(b) and (c)]. This has important implications in
SERS, since Ag substrates of the kind studied here, when illuminated
at wavelengths $\lambda< 600$ nm, could enhance the Raman signal
coming from a vibrational mode of the molecule sensitive to the
tangential electric field, as well as those sensitive to the normal
electric field (typically established as predominant according to SERS
selection rules \cite{sers_wok}).

Finally, note that using water as solvent does not introduce
significant changes in the qualitative and quantitative behavior of
the maximum local FE. In addition, we would like to point out that the
magnetic FE follows qualitatively (and almost quantitatively) the
normal electric FE.

\section{Statistical properties of the surface FE}
\label{sec_sta}

In this section we study the statistical properties of the FE's
occurring on the surface of self-affine fractal profiles with fractal
dimension and rms roughness as in the preceding section. These
properties are obtained from Monte Carlo numerical simulations results
performed as described in Sec.~\ref{sec_sform_ni}. Preliminary
results based on magnetic field calculations for weaker fractal
surfaces have been presented in Ref.~\onlinecite{jcp98}.

\subsection{Probability Density Function}
\label{sec_sta_pdf}

\begin{figure}
\epsfxsize=3.25in \epsfbox{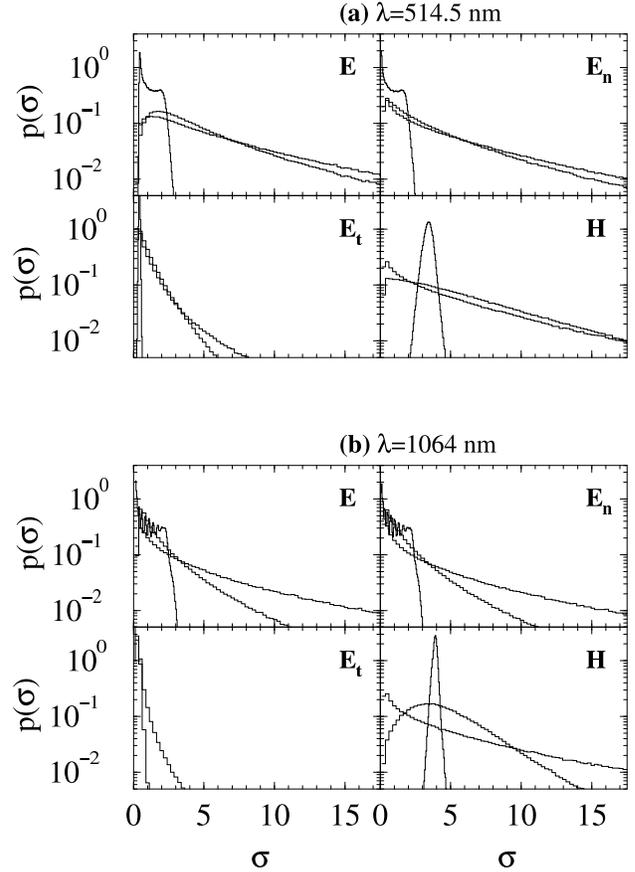}
\caption{PDF of the $p$-polarized FE factor $p(\sigma)$ for the electric
        (including both normal and tangential components) and magnetic fields
        resulting from the $p$-polarized scattering with
        $\theta_0=0^{\circ},5^{\circ},10^{\circ},\ldots ,50^{\circ}$,
        and $W=L/4\cos\theta_0$, from fractal metal surfaces consisting of
        $N_r=60$ realizations of $L=10.29 \mu$m and $N_p=1600$. Solid curve:
        $D=1.9$ and $\delta=514.5$ nm; dashed curve: $D=1.9$ and
        $\delta=102.9$ nm; and dotted curve: $D=1.2$ and $\delta=102.9$ nm.
        (a) $\lambda=514.5$ nm and (b) $\lambda=1064$ nm.}
\label{fig_pdfe}
\end{figure}

In Sec. \ref{sec_loc} we have found, by direct observation of the
calculated near-field excited in the neighborhood of the interface,
that very large fields can be excited at the surface. These enhanced
excitations are coupled to the surface by the surface roughness and,
thus, depend strongly on its properties. The question then arises as
to how probable these values are and, in turn, how the EM field
intensity is distributed over the fractal surface.  In
Fig.~\ref{fig_pdfe}, we show the PDF of the surface EM FE (including
separately tangential and normal components) for self-affine fractal
surfaces with $D=1.9$ and $\delta=$514.5 nm for two incoming
wavelengths $\lambda=$ 514.5 and 1064 nm, averaging over various
angles of incidence.  For the sake of comparison, the results for
fractal surfaces with smaller rms height $\delta=$102.9 nm, and also
with both smaller fractal dimension $D=1.2$ and rms height
$\delta=$102.9 nm are included (the latter as used in
Ref.~\onlinecite{jcp98} for magnetic FE calculations).

For the smoother surface, the resulting PDF is a narrow distribution
centered at the surface EM FE value for a flat metal surface, as
expected for such a weakly rough surface and in agreement with
Refs.~\onlinecite{oc97,jcp98}, wherein similar results were shown for the
intensity of the magnetic field (typically, $\sigma_H\sim |1+R|^2$,
$R$ being the corresponding Fresnel reflection coefficient). Recall
that the electric field components on flat metal surfaces follow
$\sigma_t\sim \cos^2\theta_0|1+R|^2$ and $\sigma_n\sim \sin^2
\theta_0|1-R|^2$, so that the PDF distributions for $D=1.2$ in
Fig.~\ref{fig_pdfe} are only significant for FE's approximately in
between the minimum and maximum expected values for the different
$\theta_0$.

For rougher surfaces, however, the surface EM no longer resembles the
flat surface result, presenting alternating dark and bright regions,
and giving rise with increasing roughness parameters to very bright
hot spots surrounded by large dark regions. The corresponding PDF
becomes wider, turning into a slowly decaying function that is maximum
at zero and exhibits a long tail for large FE values (see
Ref.~\onlinecite{jcp98} for the magnetic FE PDF for the self-affine
fractal with $D=1.9$ and $\delta=$102.9 nm). Upon comparing
Figs.~\ref{fig_pdfe} (a) and (b), it is evident that moderately large
$\sigma_t$'s become feasible at $\lambda=$514.5 nm as well as very
large $\sigma_n$'s, whereas only the $\sigma_n$'s are expected to be
intense at $\lambda=$1064 nm.

\subsection{Average and Fluctuations}
\label{sec_sta_af}

As a result of the change in the surface EM field PDF for increasing
surface roughness parameters, the moments of the distribution are also
modified. Particularly relevant are the average and the statistics of
the fluctuations, as they can also provide some information about the
global response of larger surfaces (of the order of centimeters) under
broad beam illumination.

\begin{figure}
\epsfxsize=3.25in \epsfbox{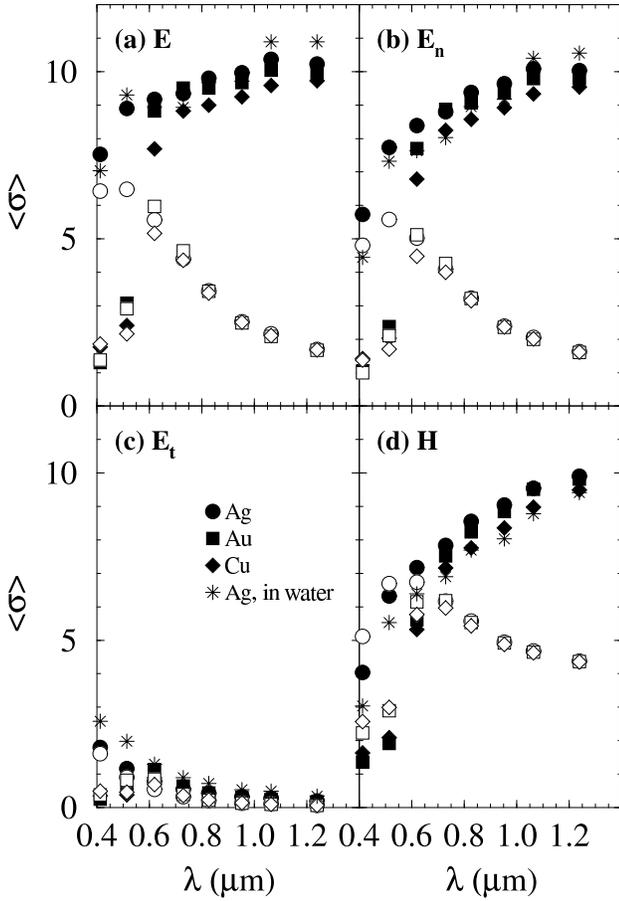}
\caption{Spectral dependence of the average surface FE $\langle\sigma\rangle$
        (excitation spectra) resulting
        from the $p$-polarized scattering with
        $\theta_0=0^{\circ},5^{\circ},10^{\circ},\ldots ,50^{\circ}$,
        and $W=L/4\cos\theta_0$, from fractal metal
        surfaces with $D=1.9$, consisting of $N_r=60$
        realizations of $L=10.29 \mu$m  and $N_p=1600$.
        (a) Electric field; (b) Electric field, normal component; (c) Electric
        field, tangential component; (d) Magnetic field.
        Circles, Ag; Squares, Au; Triangles, Cu.
        Filled symbols: $\delta=514.5$ nm; Hollow symbols: $\delta=102.9$ nm.
        Stars: water/Ag, $\delta=514.5$ nm.}
\label{fig_ave_w}
\end{figure}

In Fig.~\ref{fig_ave_w} we present the spectral dependence of the mean
FE for the same self-affine fractal surfaces whose local FE where
shown in Fig.~\ref{fig_mfe_w}. In fact, the qualitative behavior of
the mean FE does not differ substantially from that exhibited by the
maximum local FE.  Basically, there is a broad excitation spectra for
the rougher fractals (slightly narrower for the smoother fractal),
covering the visible and the NIR (at least up to $\lambda=2
\mu$m). The behavior is similar for Ag, Au, and Cu and, in all cases,
the electric field is predominantly normal to the surface.  The blue
part of the excitation spectra reveals, on the other hand, a rapid
decrease for Au and Cu associated with the onset of interband
transitions, whereas large normal FE's can still be found in that
spectral region for Ag fractals as well as an increase of the
tangential electric field upon approaching the surface plasma
wavelength (but still above the threshold wavelength of interband
transitions). The latter blue tangential electric FE increase is
slightly larger when water rather than air constitutes the propagating
medium. On the other hand, it should be emphasized that the rougher
fractal surfaces used in Fig.~\ref{fig_ave_w} give rise to an
estimated SERS FE factor $\langle {\cal G}_{EM}\approx 10^5$, in good
agreement with the phenomenological factor experimentally induced.
\cite{sers_wok}

\begin{figure}
\hspace*{1cm}\epsfxsize=2.5in \epsfbox{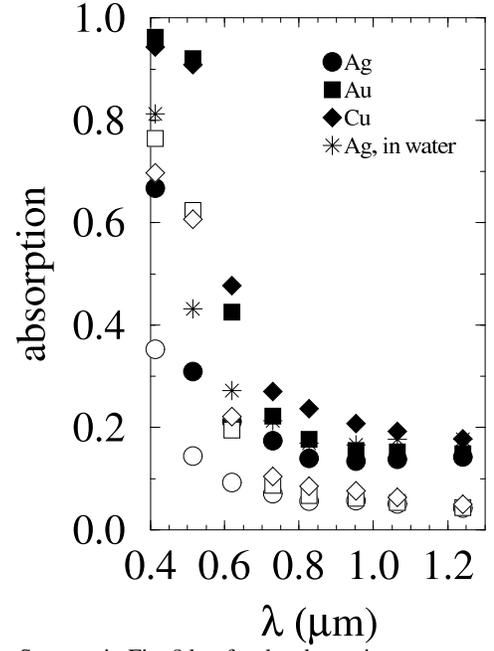}
\caption{Same as in Fig.~\protect{\ref{fig_ave_w}} but for the absorption
        spectra at normal incidence.}
\label{fig_abs_w}
\end{figure}

The absorption spectra are shown in Fig.~\ref{fig_abs_w} for the sake
of comparison. It is evident that the absorption spectrum does not
resemble the qualitative behavior of the excitation spectrum in
Fig.~\ref{fig_ave_w}. Therefore, for this kind of fractal surfaces
yielding wide excitation and absorption spectra, the maximum
absorption region as experimentally obtained from the absorption
spectra cannot be straightforwardly related to the optimum excitation
wavelength.  Furthermore, it should be emphasized that strong
absorption can even be associated with very low surface EM fields (and
thus the substrates being SERS inactive), as is the case of Au and Cu
self-affine fractals in the blue spectral region. In other substrate
configurations, nonetheless, the contrary might be the case, and
absorption bands can be used to identify excitation bands resulting in
strong surface FE (substrates becoming SERS active), as in rough
surfaces presenting surface shape plasmon resonances, such as
colloidal aggregates,\cite{sers_wok,sharep} or in gratings
diffracting into propagating SPP.\cite{spp} In summary, one has to
carefully interpret absorption spectra when using such information to
determine the appropriate SERS (or any other surface optical nonlinear
effect) excitation frequency.

\begin{figure}
\epsfxsize=3.25in \epsfbox{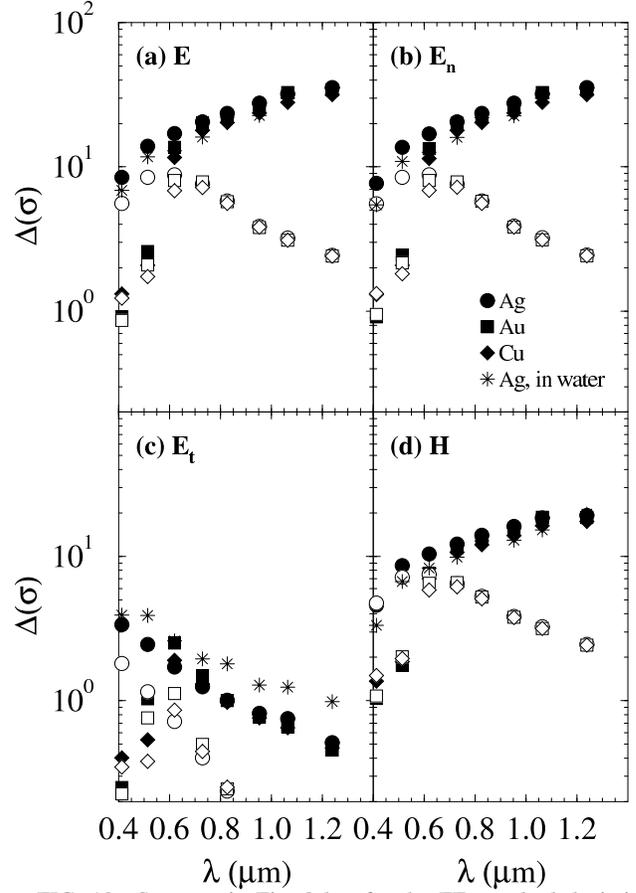}
\caption{Same as in Fig.~\protect{\ref{fig_ave_w}} but for the FE standard
        deviations $\Delta(\sigma)$.}
\label{fig_fluct_w}
\end{figure}

Finally, we show in Fig.~\ref{fig_fluct_w} the spectral dependence of
the FE fluctuations. In accordance with the previously discussed FE
PDF widening with increasing surface roughness, it is obvious that the
rougher the surface, the larger the fluctuations. And for sufficiently
rough surfaces, the fluctuations can be even larger than the average,
as seen upon comparing Fig.~\ref{fig_fluct_w} with
Fig.~\ref{fig_ave_w}. Actually, the FE fluctuations rather than the
average provide a good estimate of how probable and how bright hot
spots are. Indeed, the spectral dependence of the FE fluctuations in
Fig.~\ref{fig_fluct_w} closely follows that of the maximum local FE
shown above in Fig.~\ref{fig_mfe_w}.

\section{Conclusions}
\label{sec_con}

By means of a rigorous Green's theorem integral equation formalism, we
have studied the occurrence of strong local optical excitations (hot
spots) on self-affine fractal surfaces of Ag, Au, and Cu. The
statistics of the surface field fluctuations that produce these strong
excitations have also been studied. The formalism exploits the scalar
character of the resulting integral equations for one-dimensional
surfaces illuminated with linearly $s$- or $p$-polarized light, by
treating the problem in terms of the electric or magnetic field,
respectively. In the case of $p$ polarization, which is the relevant
one in our problem due to the SPP excitation selectivity, we have
calculated the electric field from the resulting only nonzero
component of the magnetic field and its normal derivative on the
surface.  The problem is studied numerically by means of Monte Carlo
simulations of the interaction of light with self-affine metal
fractals whose profiles were obtained from the trace of a fractional
Brownian motion. The appearance of hot spots and their statistics have
been determined for a broad spectral range of the incoming light (400
nm $<\lambda<$ 1300 nm).

We have found hot spots on self-affine fractals with fractal dimension
$D=1.9$ and rms height $\delta=$514.5 nm. These hot spots constitute
very strong and narrow (considerably narrower than half of the SPP
wavelength) surface EM field excitations, with very selective
excitation spectra (both temporally and spatially). Typically, they
give rise to local FE strongly polarized along the normal to the rough
surface. The largest ones we have found yield local SERS FE factors of
${\cal G}_{EM}\sim 10^7$, and appear for a wide range of incoming
wavelengths covering the visible and NIR up to $\lambda\approx 2
\mu$m.  Interestingly, our results reveal that weaker, tangentially
polarized hot spots can be found in Ag fractals for small excitation
wavelengths (blue or smaller).

All these features are not incompatible with the suggestion that
Anderson localization of SPP is the underlying physical mechanism
responsible for such optical excitations (e.g. the exponential decay
of the SPP transmission versus rough surface length could certainly
provide some indication of such mechanism \cite{prb97}), although no
direct evidence of this can be obtained from our scattering geometry
and calculations.

The PDF of the surface EM field for those self-affine fractals
exhibiting hot spots is a slowly decaying function with a significant
tail for large surface FE's. It differs substantially from that for
smooth surfaces for which the PDF is a narrow distribution centered at
the value of the EM field expected on a flat metal surface.

The mean FE acquires considerable values in a broad spectral
region. For the three metals considered, in most of the visible and
NIR ($\lambda<2 \mu$m) excitation regions, the component responsible
for this enhancement is the electric field normal to the surface. We
have found that, for these self-affine fractal surfaces the average
SERS FE factors are $\langle {\cal G}_{EM}\rangle\approx 10^5$.

We have also analyzed the spectral dependence of the surface FE
fluctuations.  Such fluctuations are indeed very large for the
self-affine fractals that give rise to hot spots, and present a
qualitative behavior similar to that of the maximum local FE in the
vicinity of the hot spots.  This is an interesting property that could
be used in, e.g., PSTM studies to identify samples with the potential
capability of yielding large optical excitations: even if no hot spots
are found in the region being scanned, the calculated fluctuations of
the resulting intensity map could provide a statistical account on the
probability of finding hot spots (simpler than calculating the total
PDF for which much more data from a larger scanning area would be
required).

With regard to the quantitative aspects of SERS, the maximum local
enhancement factors are still below those that could be deduced from
experimental works on single molecule detection \cite{nie97,kneipp97}
and from approximate theoretical calculations.
\cite{sharep,sha96} This, however, is not entirely surprising, given
the differences in the type of SERS substrates being considered. In
fact, the typical SERS spectroscopy enhancement factors are fairly
similar to those found in this work.

We would like to emphasize that our formulation is exact within the
classical EM framework, at least as far as the linear (direct) field
enhancement factor is concerned. Thus, our calculations should provide
a truthful picture of the linear optical response of self-affine
fractal metal substrates. Further work is of course needed to test the
perhaps naive assumption that the enhancement factor at the
Raman-shifted frequency is identical to that obtained at the
excitation frequency. Also, more work is required to study the effects
of lower scaling cutoffs in the generation of the fractal surfaces, as
these spatial frequencies might contribute to build up the enhancement
factors.\cite{pre_fe}

Finally, we mention that work involving rigorous calculations of the
kind presented here is also in progress for the study of the field
enhancements produced on self-similar substrates, such as those found
in colloidal aggregates,\cite{jcis94} and for the study surfaces
covered by a monolayer of Raman active molecules (Langmuir-Blodgett
films).

\acknowledgements

This work was supported by the Spanish Direcci{\'o}n General de Ense\~nanza
Superior e Investigaci{\'o}n Cient{\'\i}fica y T{\'e}cnica, through grant
No. PB97-1221. We also thank the Mexican-Spanish CONACYT-CSIC program for
partial travel support.


\end{document}